\documentclass{article}

\usepackage[
backend=biber,
style=numeric-comp,
sorting=nty
]{biblatex}
\usepackage[margin=1in]{geometry}
\usepackage[utf8]{inputenc}
\usepackage{csquotes}
\usepackage{amsfonts,amsthm,amsmath,amssymb,accents,mathtools,physics} 
\usepackage{mathrsfs,amsbsy} 
\usepackage{enumitem}
\usepackage{subfig}
\usepackage{graphicx}
\usepackage{float}
\usepackage{xcolor}
\usepackage{algorithm}
\usepackage[noend]{algpseudocode}
\usepackage{hyperref}
\usepackage{authblk} 
\usepackage{xeCJK} 

\usepackage{textcomp} 
\usepackage{algorithm}
\usepackage{algorithmicx}
\usepackage{relsize}
\usepackage{bm}
\usepackage{bbm}
\usepackage{dsfont}

\numberwithin{equation}{section}

\usepackage{yhmath}
 \usepackage{booktabs} 

\theoremstyle{definition}

\def\urm#1{\scriptstyle{\text{\textrm{\textmd{\textup{#1}}}}}}

\hypersetup{
    colorlinks=true,
    linkcolor=blue,
    citecolor=blue,
    filecolor=blue,      
    urlcolor=blue,
}

\title{The Pseudospectral Method for the Dirac Equation with Confining Potential}
\author[1]{Dengshan Liu}
\author[1]{Huihui Xie}
\author[1]{Pengxiang Du}
\author[1]{Jian Li \thanks{Corresponding author: jianli@jlu.edu.cn}}
\author[2,3]{Tomoya Naito}
\affil[1]{College of Physics, Jilin University, Changchun 130012, China.}
\affil[2]{Department of Nuclear Engineering and Management, Graduate School of Engineering, The University of Tokyo, Tokyo 113-8656, Japan}
\affil[3]{RIKEN Center for Interdisciplinary Theoretical and Mathematical Sciences (iTHEMS), Wako 351-0198, Japan}
\date{}

\begin{document}

\maketitle


\begin{abstract}
\noindent We observe that solving the Dirac equation for confined potentials using the generalized pseudospectral (GPS) method leads to deteriorating convergence of energy eigenvalues and highly oscillatory in wave functions as the confinement radius decreases. It is found that this issue stems from the first-order differentiation formulation employed in GPS method.  
Motivated by this insight, we adopt the kinetically-balanced generalized pseudospectral method, which incorporates the kinetically-balanced condition into the GPS method.  
Numerical results demonstrate that the mono-kinetically-balanced generalized pseudospectral (MKB-GPS) method yields converged energy eigenvalues and generates smooth, continuous wave functions.  
This is the first application of the MKB-GPS method to confined potentials, and its effectiveness is validated for small confinement radii.
\end{abstract}



\section{Introduction}

\setcounter{equation}{0}
\par
The Dirac equation, serving as the relativistic counterpart of the Schr\"{o}dinger equation in quantum mechanics,
is pivotal for describing relativistic systems composed of spin-$ 1/2 $ particles.
This equation comprises a set of coupled first-order differential equations involving both large and small components of the wave function.
Since only a limited number of potential forms permit analytical solutions,
numerical approaches are, generally, required to obtain 
the eigenenergies and wave functions of the Dirac equation.
In recent decades,
precise solutions of the Dirac equation have been paid attention to
driven by rapid advances in high-precision measurement techniques for atomic and molecular systems.
Notably, the accurate treatment of relativistic single-electron systems governed by the Dirac equation provides a foundational framework for investigating the structural and dynamical properties of complex quantum systems~\cite{SAMPSON2009111, KOZLOV2015199, ZATSARINNY2016287, FRITZSCHE20191}.
\par
The relativistic Dirac equation for the one-electron system in the Hartree atomic units
($ 4 \pi \epsilon_0 = \hbar = m_e = e^2 = 1 $)
reads 
\begin{equation}
  \label{SDEd}
  \left[
    c \bm{\alpha} \cdot \bm{p} + \beta c^2 + V \left( \bm{r} \right)
  \right]
  \psi \left( \bm{r} \right)
  =
  E \psi \left( \bm{r} \right),  
\end{equation}
where the eigenenergy $ E $ includes both the electron binding energy $ \varepsilon $ and the rest mass of electron $ c^2 $, i.e.,
$ E = \varepsilon + c^2 $.
In this paper, we assume the spherical symmetry for the electrostatic potential $ V \left( \bm{r} \right) $,
i.e., $ V \left( \bm{r} \right) = V \left( r \right) $.
Then, the electron wave functions can be written as
\begin{equation}
  \label{SDEd-init}
  \psi_{n \kappa m} \left( \bm{r} \right)
  =
  \frac{1}{r}
  \begin{pmatrix}
    P_{n \kappa}   \left( r \right) \chi_{\kappa  m} \left( \theta, \varphi \right) \\
    i Q_{n \kappa} \left( r \right) \chi_{-\kappa m} \left( \theta, \varphi \right)
  \end{pmatrix} ,  
\end{equation}
where $ P_{n \kappa} \left( r \right) $ and $ Q_{n \kappa} \left( r \right) $ are the large and small components of the radial wave functions, respectively.
The angular parts $ \chi_{\pm \kappa m} \left( \theta, \varphi \right) $ are eigenstates of
$ \bm{L}^2 $, $ \bm{S}^2 $, $ \bm{J}^2 $, and $ J_z $
with the eigenvalues being $ l \left( l + 1 \right) $, $ 3/4 $, $ j \left( j + 1 \right) $, and $ m $,
and $ \chi_{\pm \kappa m} \left( \theta, \varphi \right) $ is two-component spinor.
The Dirac quantum number $ \kappa $ is connected with the total and orbital angular momenta $ j $ and $ l $ by
\begin{equation}
  \label{kappa}
  \kappa
  =
  l \left( l + 1 \right)
  -
  j \left( j + 1 \right)
  -
  \frac{1}{4}.  
\end{equation}
After separating out the angular component of the wave function, the radial Dirac equation for the one-electron system is given by
\begin{equation}
  \label{eq14}
  \begin{pmatrix}
    V \left( r \right) + c^2  & - c \left( \frac{d}{dr} - \frac{\kappa}{r} \right) \\
    c \left( \frac{d}{dr} + \frac{\kappa}{r} \right) & V \left( r \right) - c^2
  \end{pmatrix}
  \begin{pmatrix}
    P_{n \kappa} \left( r \right) \\
    Q_{n \kappa} \left( r \right)
  \end{pmatrix}
  =
  E
  \begin{pmatrix}
    P_{n \kappa} \left( r \right) \\
    Q_{n \kappa} \left( r \right)
  \end{pmatrix}.
\end{equation}
The radial Dirac spinors $ P \left( r \right) $ and $ Q \left( r \right) $ for bound states are normalized by
\begin{equation}
  \int_0^{\infty}
  \left[
    P_{n \kappa}^2 \left( r \right)
    +
    Q_{n \kappa}^2 \left( r \right)
  \right]
  \, dr
  =
  1,
\end{equation}
and fulfills the Dirichlet boundary condition
\begin{equation}  
  \label{eq16}
  \psi \left( 0 \right)
  =
  \psi \left( \infty \right)
  =
  0.
\end{equation}
Unlike the nonrelativistic Schr\"{o}dinger equation,
variational calculation of the radial Dirac equation generally encounters the issue of spurious states when approximating the spectrum of the Dirac Hamiltonian with a sequence of finite-dimensional basis sets~\cite{Tupitsyn2008183,Lewin2010}.
These spurious states arise due to the derivative formulations employed in numerical methods~\cite{zhaobin2016,Szafran2019,Salomonson1989,MULLER1998}.
For instance, in the finite-difference method, the use of the combination of forward and backward scheme of the numerical derivative can eliminate spurious states introduced by central-difference representation of the first-order derivative~\cite{Salomonson1989,zhang2022}.
\par
An alternative method to avoid such an issue is to introduce an extra Wilson term into the Hamiltonian,
which helps distinguish physical states from spurious solutions with enhancing the accuracy of the finite-difference approach~\cite{Tanimura2015,Fang2020}.
If one expands wave functions and operators using a basis, spurious states can be eliminated by introducing the kinetically-balanced condition~\cite{Shabaev2004130405}.
This condition imposes a specific constraint between the large and small components of the radial wave functions and ensures that the Dirac equation reduces to the Schr\"{o}dinger equation in the non-relativistic limit.
Furthermore, the condition can also be effectively applied to suppress spurious states in the pseudospectral method~\cite{Jiao2021},
which is a global method of a discrete variable representation that transforms differential equations into matrix eigenvalue problems, featuring exponential convergence and the advantage of avoiding the computation of complicated potential matrix elements.
\par
On top of such a problem related to spurious states,
a new challenge is introduced to the pseudospectral method for numerically solving the Dirac equation
under the confinement potential within a relatively small spatial region
\begin{equation}
  V \left( r \right)
  =
  \begin{cases}
    - \frac{Z}{r} & r < R_{\urm{max}}, \\
    \infty        & r > R_{\urm{max}}.
  \end{cases}
\end{equation}
That is high-frequency oscillation emerges in one component of the wave functions.
The oscillation worsens the convergence of energy eigenvalues, and crucially, this phenomenon persists across all obtained eigenstates.
Such behavior occurs only when the cutoff radius $ R_{\urm{max}} $ is relatively small;
as $ R_{\urm{max}} $ approaches infinity, the numerical results approach those of standard hydrogen-like ions~\cite{Jiao2021,ZhuLin2020,wangcx}.
However, we will demonstrate that introducing a kinetically-balanced condition can effectively resolve such issues,
yielding smooth and continuous wave functions along with highly convergent energy eigenvalues.
The key insight lies in incorporating a second-order differential operator into the Dirac Hamiltonian through the kinetically-balanced condition.
Therefore, this work will systematically analyzes how a kinetically-balanced condition addresses these numerical challenges and verifies the computational results of the kinetically-balanced generalized pseudospectral method through numerical experiments.
\par
The rest of the paper is organized as follows.
Section~\ref{sec:GPS} will present the generalized pseudospectral (GPS) method for the Dirac equation,
and demonstrates the results of solving the Dirac equation with the confining potential using this method.
Section~\ref{sec:MTI} will introduce two sets of equations to be satisfied for the Dirac equation.
The mono-kinetically-balanced generalized pseudospectral (MKB-GPS) method will be described in Section~\ref{sec:GPM},
along with its numerical performance.
Finally, some conclusions will be drawn in Section~\ref{sec:con}.
%


\section{Generalized pseudospectral method}
\label{sec:GPS}
\par
The generalized pseudospectral method was first introduced by Yao and Chu~\cite{YAO1993}
to efficiently solve the radial Schr\"{o}dinger and Dirac equations in discrete variable representation.
Details about the GPS method are available elsewhere~\cite{Jiao2021,CHU2004,Telnov2013,Canuto2006,ZhuLin2020},
and here we only give a brief summary of it.
The radial variable $ r \in \left[ 0, R_{\urm{max}} \right] $ is first mapped onto a finite interval of $ x \in \left[-1, 1 \right] $ by utilizing a rational mapping function
\begin{equation}
  \label{eq21}
  r
  \equiv
  f \left( x \right)
  =
  L
  \frac{1 + x}{1 - x + \alpha}
\end{equation}
with $ \alpha = 2L/R_{\urm{max}} $,
where $ L $ is an adjustable mapping parameter.
The radial wave function $ \psi \left( r \right) $ is then transformed to
\begin{equation}
  \label{eq22}
  \phi \left( x \right)
  =
  \sqrt{f' \left( x \right)}
  \psi \left( r \right)
\end{equation}
\par
In the Dirac equation, the radial wave functions defined in Eq.~\eqref{SDEd-init} have two components as 
\begin{equation}
  \psi \left( r \right)
  =
  \begin{pmatrix}
    P \left( r \right) \\
    Q \left( r \right)
  \end{pmatrix},
\end{equation}
in which $ P \left( r \right) $ and $ Q \left( r \right) $ are, respectively, the large and small components;
accordingly $ \phi $ also has two components as
\begin{equation}
  \label{eq23}
  \phi \left( x \right)
  =
  \begin{pmatrix}
    p \left( x \right) \\
    q \left( x \right)
  \end{pmatrix}.
\end{equation}
After several algebraic manipulations,
one gets the new coupled first-order differential equation with respect to the variable $ x $ as 
\begin{equation}
  \label{eq25}
  h_{\urm{D}} \left( x \right)
  \phi \left( x \right)
  =
  E \phi \left( x \right)
\end{equation}
with 
\begin{equation}
  \label{eq26}
  h_{\urm{D}} \left( x \right)
  =
  \begin{pmatrix}
    V \left( f \left( x^2 \right) \right) + c^2 & c \kappa / f \left( x \right) \\
    c \kappa / f \left( x \right) & V \left( f \left( x^2 \right) \right) - c^2
  \end{pmatrix}
  +
  \begin{pmatrix}
    0 & -c \\
    c &  0
  \end{pmatrix}
  \frac{1}{\sqrt{f' \left( x \right)}}
  \frac{d}{dx}
  \frac{1}{\sqrt{f' \left( x \right)}}.
\end{equation}
The Dirichlet boundary conditions of Eq.~\eqref{eq16} corresponds to 
\begin{equation}
  \phi \left( -1 \right)
  =
  \phi \left( 1 \right)
  =
  0.
\end{equation}
The two-component unknown function $ \phi \left( x \right) $ is approximated by two $ N $th-order polynomials
$ \phi_N \left( x \right) $ expressed in terms of superposition of cardinal functions
\begin{equation}
  \label{eq30}
  \phi_N \left( x \right)
  =
  \sum_{j = 0}^N
  g_j \left( x \right)
  \phi \left( x_j \right)
  =
  \sum_{j = 0}^N
  \left [
    \begin{pmatrix}
      g_j \left( x \right) \\
      0
    \end{pmatrix}
    p \left( x_j \right)
    +
    \begin{pmatrix}
      0 \\
      g_j \left( x \right)
    \end{pmatrix}
    q \left( x_j \right)
  \right],
\end{equation}
where $ x_j $ are the collocation points with $ x_0 = -1 $ and $ x_N = 1 $,
and $ x_1 $, $ x_2 $, \ldots, and $ x_{N - 1} $ are roots of the first derivative of the Lagrange polynomial $ P_N \left( x \right) $.
The cardinal functions $ g_j \left( x \right) $ is constructed using Lagrange polynomials as
\begin{equation}
  \label{eq:def_g}
  g_j \left( x \right)
  =
  \frac{1}{N \left( N + 1 \right)}
  \frac{\left( 1 - x^2 \right)}{\left( x_j - x \right)}
  \frac{P'_N \left( x \right)}{P_N \left( x_j \right)},
\end{equation}
which fulfills the condition $ g_j \left( x_i \right) = \delta_{ij} $. 
Therefore, the equality $ \phi_N \left( x \right) = \phi \left( x \right) $ exactly holds at $ x = x_j $
and thus we can assume that $ \phi_N $ is an approximated function of $ \phi $.
The substitution of Eq.~\eqref{eq30} into Eq.~\eqref{eq25} leads to the discretized formulation
\begin{equation}
  \label{eq38-1}
  \sum_{j = 0}^N
  \left[
    h_{\urm{D}} \left( x \right)
    g_j \left( x \right)
  \right]
  \phi \left( x_j \right)
  =
  E
  \sum_{j = 0}^N
  g_j \left( x \right)
  \phi \left( x_j \right).
\end{equation}
Based on the property of the basis functions $ g_j \left( x_i \right) = \delta_{ij} $,
Eq~\eqref{eq38-1} can be further discretized, leading to the standard eigenvalue problem:
\begin{equation}  
  \label{eq39-1}  
  \left[ \mathbf{h}_{\urm{D}} \right]  
  \left[ \mathbf{A} \right]  
  =  
  \left[ \mathbf{E} \right]  
  \left[ \mathbf{A} \right],
\end{equation}
where the matrix element is
\begin{equation}
  \label{eq39-2}
  \left( h_{\urm{D}} \right)_{ij}
  =
  \left.
  h_{\urm{D}} \left( x \right)
  g_j \left( x \right)
  \right|_{x = x_i}.
\end{equation}
In the calculation of the matrix elements $ \left( h_{\urm{D}} \right)_{ij} $,
it is also necessary to evaluate the first-derivative of the function at the collocation points.
An arbitrary function $ u \left( x \right) $ can be expressed as a superposition of the basis functions $ g_j \left( x \right) $
\begin{equation}
  \label{eq39-3}
  u \left( x \right)
  =
  \sum_{j = 0}^N
  g_j \left( x \right)
  u \left( x_j \right).
\end{equation}
Then, its first-order derivative at arbitrary collocation point has the form
\begin{equation}
  \label{eq39-4}
  \left.
    \frac{d}{dx}
    u \left( x \right)  
  \right|_{x = x_i}
  =
  \sum_{j = 0}^N
  g'_j \left( x_i \right)
  u \left( x_j \right)
  =
  \sum_{j = 0}^N
  \left( d_1 \right)_{ij}
  \frac{P_N \left( x_i \right)}{P_N \left( x_j \right)}
  u \left( x_j \right),
\end{equation}
where
\begin{equation}  
  \label{eq39-5}  
  \left( d_1 \right)_{ij}  
  =  
  \begin{cases}    
    - \frac{1}{x_i - x_j} & \text{for $ i \ne j $}, \\    
    0                     & \text{for $ i = j \in \left[ 1, N-1 \right] $}, \\    
    - \frac{N \left( N + 1 \right)}{4} & \text{for $ i = j = 0 $}, \\    
    \frac{N \left( N + 1 \right)}{4}   & \text{for $ i = j = N $}.  
  \end{cases}
\end{equation}
Therefore, the Hamiltonian matrix element in Eq.~\eqref{eq39-1} reads
\begin{equation}
  \label{eq39-6}
  \left( h_{\urm{D}} \right)_{ij}
  =
  \begin{pmatrix}
    V \left( f \left( x_i \right) \right) + c^2 & c \kappa / f \left( x_i \right) \\
    c \kappa / f \left( x_i \right) & V \left( f \left( x_i \right) \right) - c^2
  \end{pmatrix}
  \delta_{ij}
  +
  \begin{pmatrix}
    0 & -c \\
    c &  0
  \end{pmatrix}
  \frac{1}{\sqrt{f' \left( x_i \right)}}
  \left( d_1 \right)_{ij}
  \frac{1}{\sqrt{f' \left( x_j \right)}}
  \frac{P_N \left( x_i \right)}{P_N \left( x_j \right)} .
\end{equation}
Currently, the Hamiltonian matrix $ \left[ \mathbf{h}_{\urm{D}} \right]$ is nonsymmetric
because $ \left( h_{\urm{D}} \right)_{ij} \ne \left( h_{\urm{D}} \right)_{ji} $ holds.
One convenient way to convert it to be symmetric is to utilize the transformation
\begin{align}
  \label{eq39-7}
  \left( H_{\urm{D}} \right)_{ij}
  & =
    \left( h_{\urm{D}} \right)_{ij}
    \frac{P_{\urm{N}} \left( x_j \right)}{P_N \left( x_i \right)} \\
  & =
    \begin{pmatrix}
      V \left( f \left( x_i \right) \right) + c^2 & c \kappa / f \left( x_i \right) \\
      c \kappa / f \left( x_i \right) & V \left( f \left( x_i \right) \right) - c^2
    \end{pmatrix}
                                        \delta_{ij}
                                        +
                                        \begin{pmatrix}
                                          0 & -c \\
                                          c &  0
                                        \end{pmatrix}
                                              \frac{1}{\sqrt{f' \left( x_i \right)}}
                                              \left( d_1 \right)_{ij}
                                              \frac{1}{\sqrt{f' \left( x_j \right)}}.
\end{align}
and the corresponding eigenvector reads
\begin{equation}
  \label{eq39-8}
  A_j  
  =  
  \frac{\phi \left( x_j \right)}{P_N \left( x_j \right)}  
  =  
  \frac{\sqrt{f' \left( x_j \right)} \psi \left( f \left( x_j \right) \right)}{P_N \left( x_j \right)}.
\end{equation}
The final standard symmetric eigenvalue problem is expressed as
\begin{equation}  
  \label{eq39-9}  
  \left[ \mathbf{H}_{\urm{D}} \right]  
  \left[ \mathbf{A} \right]  
  =  
  \left[ \mathbf{E} \right]  
  \left[ \mathbf{A} \right].
\end{equation}
Here $ \left[ \mathbf{A} \right] $ is a square matrix storing all eigenvectors columnwise,
and $ \left[ \mathbf{E} \right] $ is a diagonal matrix containing the corresponding eigenvalues on its diagonal.
Both have the same dimension as $ \left[ \mathbf{H}_{\urm{D}} \right] $.
\par
\begin{figure}
  \centering
  \includegraphics[width=1.0\textwidth]{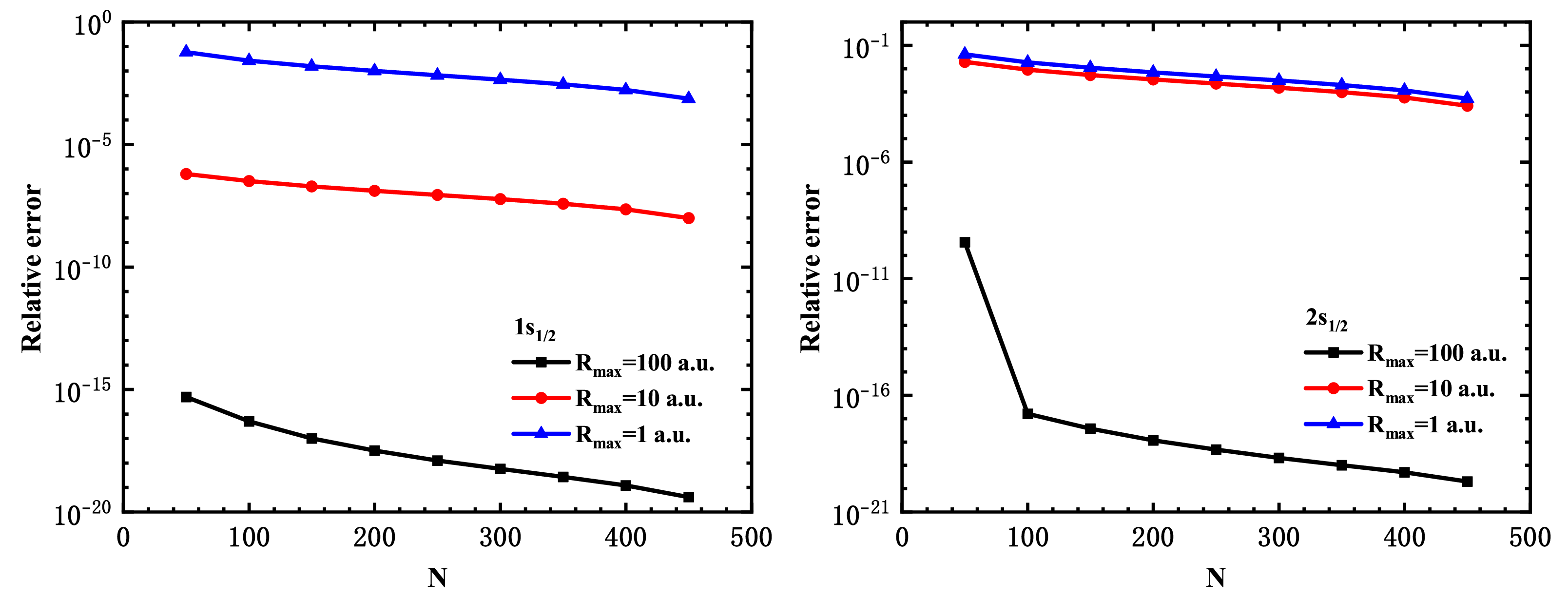}
  \caption{Relative errors of the bound-state energies of the hydrogen atom calculated by GPS method with the different confinement radii $ R_{\urm{max}} $.}
  \label{fig:figure1}
\end{figure}
\begin{figure}
  \centering
  \includegraphics[width=1.0\textwidth]{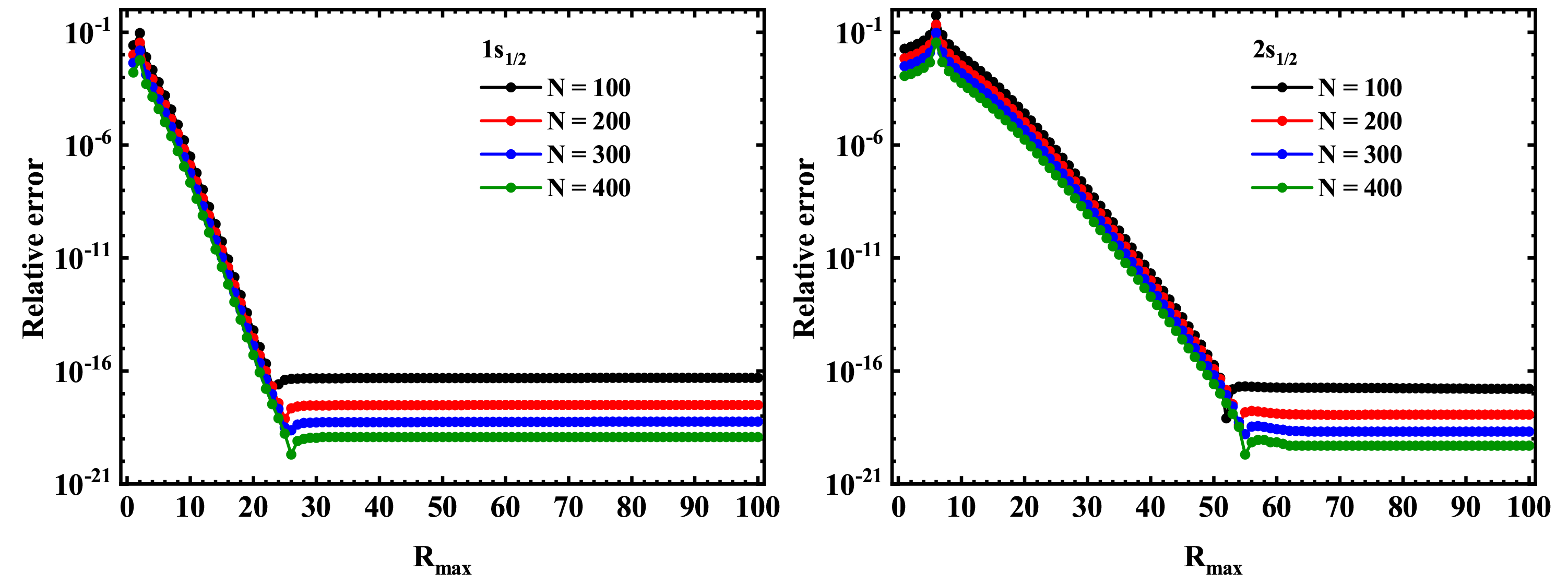}
  \caption{Relative errors of the bound-state energies of the hydrogen atom calculated by GPS method with the different $ N $.}
  \label{fig:figure1_1}
\end{figure}
\begin{figure}
  \centering
  \includegraphics[width=1.0\textwidth]{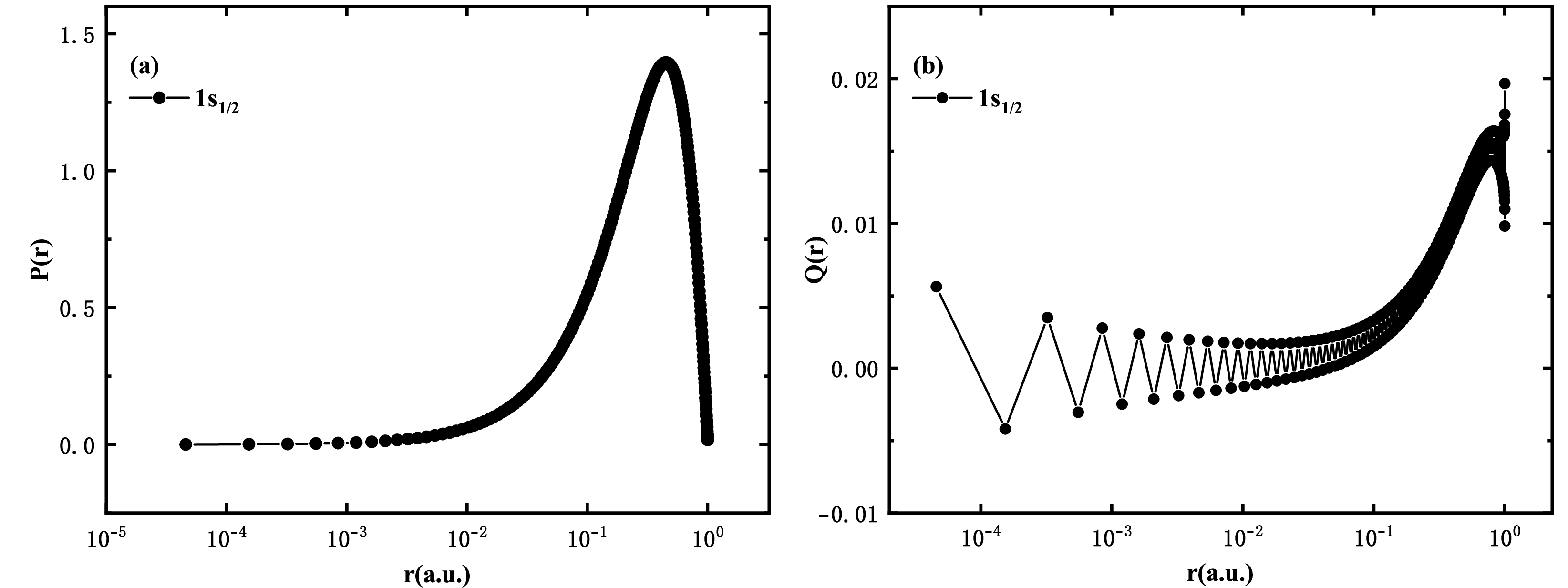}
  \caption{The radial wave function of the $ 1s_{1/2} $ state of the hydrogen atom calculated by GPS method with the confinement radius $ R_{\urm{max}} = 1 \, \mathrm{a.u} $.
    (a) Large component and (b) small component.}
  \label{fig:figure2}
\end{figure}
\par
Figures~\ref{fig:figure1} and \ref{fig:figure1_1} display the relative errors of bound-state energies for the hydrogen atom, computed using the GPS method.
The reference energies are obtained using the GPS method with $ N = 500 $.
To model hydrogen atoms under extreme pressure conditions,
it is necessary to set the confinement radius $ R_{\urm{max}} $ to a relatively small value, e.g., $ R_{\urm{max}} = 1 \, \mathrm{a.u} $.
It can be seen from Fig.~\ref{fig:figure1} that for both the $ 1s_{1/2} $ and $ 2s_{1/2} $ states, the relative errors for different confinement radii $ R_{\urm{max}} $ decrease as the number of grid points $ N $ increases, with a particularly pronounced decrease when  $ R_{\urm{max}} = 100 \, \mathrm{a.u} $.
As can be seen more clearly from Fig.~\ref{fig:figure1_1},
for the $ 1s_{1/2} $ state, when the confinement radius $ R_{\urm{max}} > 25 \, \mathrm{a.u.} $,
the relative error is very small, on the order of $ 10^{-16} $;
for the $ 2s_{1/2} $ state, the relative error is also very small for $ R_{\urm{max}} > 50 \, \mathrm{a.u} $.
However, when the confinement radius is reduced to $ 1 \, \mathrm{a.u.} $, the relative error becomes significant.
Moreover, Fig.~\ref{fig:figure2} reveals that under this confinement condition ($ R_{\urm{max}} = 1 \, \mathrm{a.u.} $),
high-frequency oscillation emerges in the small component of the wave functions,
violating the requirement for smooth and continuous wave functions.
On the other hand, the large component shown in Fig.~\ref{fig:figure2} (a) appears to be a smooth function and thus presents no apparent issues.
In addition, when solving the Schr\"{o}dinger equation under the same conditions, this problem does not arise.
The consistency between the large and small components is discussed in Section~\ref{sec:MTI}.
%


\section{Two sets of equations to be satisfied for the Dirac equation}

\label{sec:MTI}
\par
We begin with the radial Dirac equation to analyze the numerical challenges encountered in the generalized pseudospectral method.
From Eq.~\eqref{eq14}, we obtain
\begin{subequations}
  \label{eq40}
  \begin{align}
    \left[
    V \left( r \right)
    +
    c^2
    -
    E \right]
    P \left( r \right)
    -
    c
    \left(
    \frac{d}{dr}
    -
    \frac{\kappa}{r}
    \right)
    Q \left( r \right)
    & =
      0, \\
    c
    \left(
    \frac{d}{dr}
    +
    \frac{\kappa}{r}
    \right)
    P \left( r \right)
    +
    \left [
    V \left( r \right)
    -
    c^2
    -
    E
    \right]
    Q \left( r \right)
    & =
      0;
  \end{align}
\end{subequations}
then, we obtain the relation between $ P $ and $ Q $ as 
\begin{subequations}
  \begin{align}
    P \left( r \right)
    & =
      \frac{c}{V \left( r \right) + c^2 - E}
      \left(
      \frac{d}{dr}
      -
      \frac{\kappa}{r}
      \right)
      Q \left( r \right), 
      \label{eq:41} \\
    Q \left( r \right)
    & =
      \frac{c}{c^2 + E - V \left( r \right)}
      \left(
      \frac{d}{dr}
      +
      \frac{\kappa}{r}
      \right)
      P \left( r \right).
      \label{eq:42}
  \end{align}
\end{subequations}
Hereinafter, we refer to Eqs.~\eqref{eq:41} and \eqref{eq:42} as the first set of equations to be satisfied.
The substitution of Eq.~\eqref{eq:42} into Eq.~\eqref{eq:41} simply yields the equation
\begin{equation}
  \label{eq:43}
  \left[
    V \left( r \right)
    +
    c^2
    -
    E
  \right]
  P \left( r \right)
  -
  \frac{c^2}{E + c^2 - V \left( r \right)}
  \left(
    \frac{d}{dr}
    -
    \frac{\kappa}{r}
  \right)
  \left(
    \frac{d}{dr}
    +
    \frac{\kappa}{r}
  \right)
  P \left( r \right)
  =
  0,
\end{equation}
which is equivalent to
\begin{subequations}
  \begin{align}
    \frac{d^2}{dr^2}
    P \left( r \right)
    -
    \frac{\kappa \left( \kappa + 1 \right)}{r^2}
    P \left( r \right)
    & =
      \frac{\left[ V \left( r \right) + c^2 - E \right] \left[ c^2 + E - V \left( r \right) \right]}{c^2}
      P \left( r \right)
      \notag \\
    & =
      \left\{
      -
      \frac{\left[ V \left( r \right) - E \right]^2}{c^2}
      +
      c^2
      \right\}
      P \left( r \right).
      \label{eq:44}
  \end{align}
  Similarly, substitution of Eq.~\eqref{eq:41} into Eq.~\eqref{eq:42} yields
  \begin{align}
    \frac{d^2}{dr^2}
    Q \left( r \right)
    -
    \frac{\kappa \left( \kappa - 1 \right)}{r^2}
    Q \left( r \right)
    & =
      \frac{\left[ c^2 + E -  V \left( r \right) \right] \left[ c^2 + V \left( r \right) - E \right]}{c^2}
      Q \left( r \right)
      \notag \\
    & =
      \left\{
      -
      \frac{\left[ V \left( r \right) - E \right]^2}{c^2}
      +
      c^2
      \right\}
      Q \left( r \right).
      \label{eq:45}
  \end{align}
\end{subequations}
Hereinafter, we refer to Eqs.~\eqref{eq:44} and \eqref{eq:45} as the second set of equations to be satisfied.
\begin{figure}
  \centering
  \includegraphics[width=1.0\textwidth]{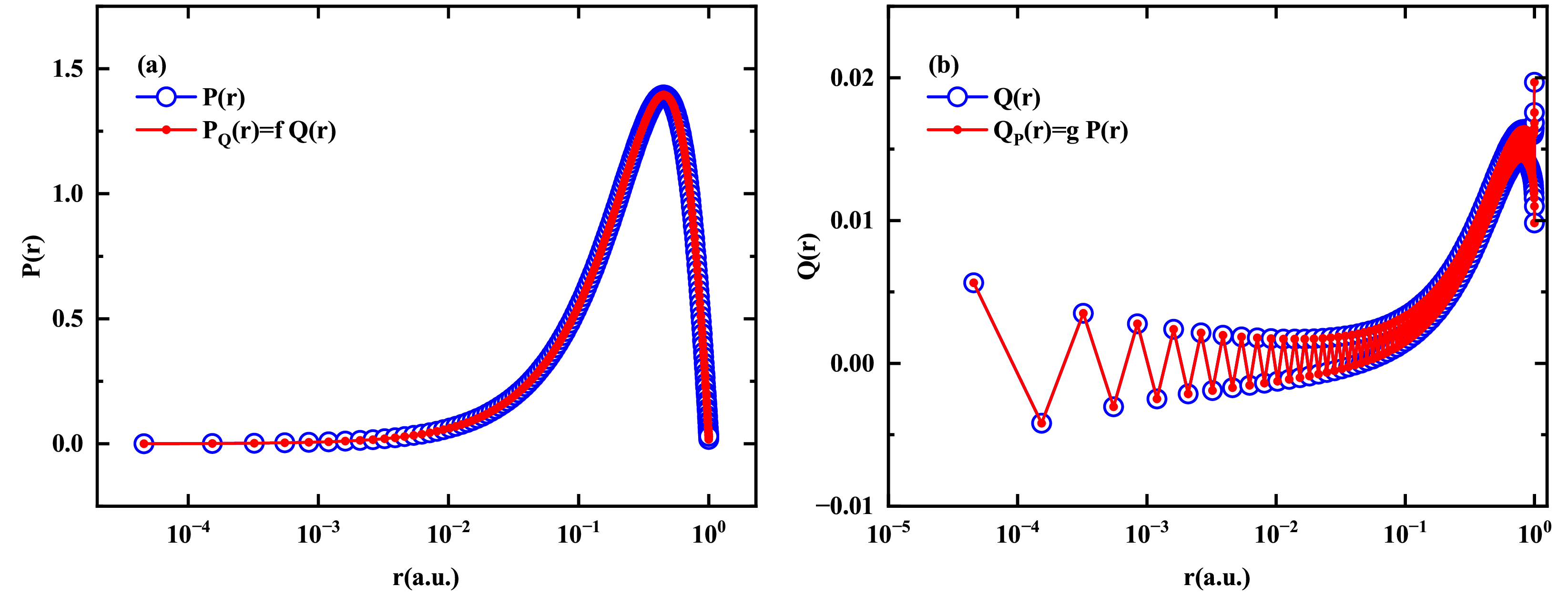}
  \caption{Radial wave functions of the $1s_{1/2}$ state calculated by GPS method (blue hollow circle and line)
    and Eqs.~\eqref{eq:41} and \eqref{eq:42} (red solid circle and line).}
  \label{fig:figure3}
\end{figure}
\par
First, we examine whether the large and small components of the wave function shown in Fig.~\ref{fig:figure2}
satisfy the first set of equations [Eqs.~\eqref{eq:41} and \eqref{eq:42}].
The corresponding results are shown in Fig.~\ref{fig:figure3},
where the large component $ P \left( r \right) $ and the small component $ Q \left( r \right) $ of the wave function of the $1s_{1/2}$ state obtained by GPS method are represented by blue hollow circle and line.
On top of it,
$ P \left( r \right) $ obtained by the combination of $ Q \left( r \right) $ and Eq.~\eqref{eq:41}
and 
$ Q \left( r \right) $ by $ P \left( r \right) $ and Eq.~\eqref{eq:42}
are plotted as red solid circle and line.
The former and latter are, respectively, indicated as
``$ P_Q \left( r \right) = f Q \left( r \right) $''
and 
``$ Q_P \left( r \right) = g P \left( r \right) $''
and 
the derivatives of the wave function are computed using Eq.~\eqref{eq39-4}.
Even though the small component of the wave function $ Q \left( r \right) $ exhibits rapid oscillations,
$ P_Q \left( r \right) $ obtained by $ Q \left( r \right) $ and Eq.~\eqref{eq:41} unexpectedly
give a smooth and continuous wave function that agrees well with the large component $ P \left( r \right) $ obtained by the GPS method
as shown in Fig.~\ref{fig:figure3}.
Similarly,
$ Q_P \left( r \right) $ successfully reproduces the rapidly oscillating small component $ Q \left( r \right) $.
Therefore, both the large and the small components of the wave functions satisfy the first set of equations to be satisfied with excellent agreement.
\par
The first equations set [Eqs.~\eqref{eq:41} and \eqref{eq:42}] 
incorporates first-order differential operations on the wave functions.
Nevertheless, the oscillatory behavior unexpectedly vanishes
when the numerical derivative [Eq.~\eqref{eq39-4}] is applied to perform differentiation on the rapidly oscillating small component to derive the large component.
Conversely, first-order differentiation of the large component generates oscillatory features in the resulting small component.
These observations reveal that the fundamental issue stems from the first-order differentiation formulation within the GPS method.
\begin{figure}
  \centering
  \includegraphics[width=1.0\textwidth]{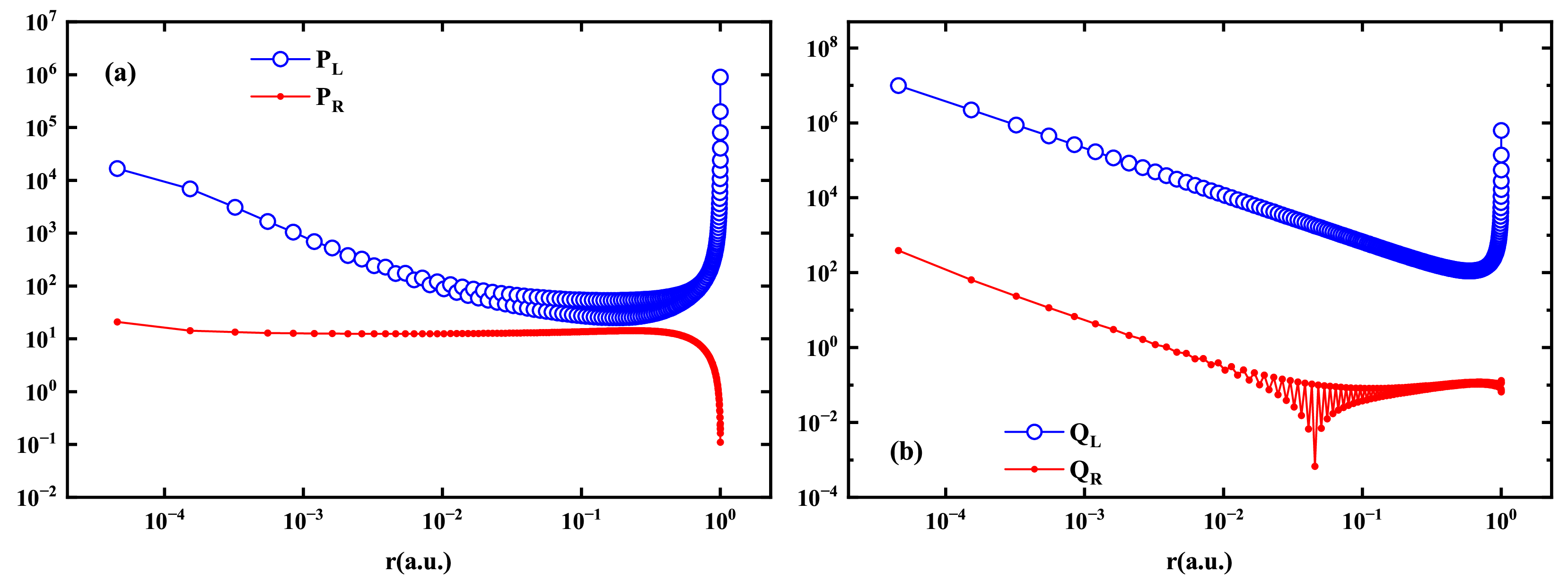}
  \caption{Functions on the left- and right-hand sides of the second set of equations to be satisfied [Eqs.~\eqref{eq:44} and \eqref{eq:45}].}
  \label{fig:figure4}
\end{figure}
\par
Therefore, to resolve the current issue,
we employ the second set of equations to be satisfied [Eqs.~\eqref{eq:44} and \eqref{eq:45}]
to examine the wave function in Fig.~\ref{fig:figure2}, with the corresponding computational results presented in Fig.~\ref{fig:figure4}.
In Fig.~\ref{fig:figure4} (a), $ P_L $ and $ P_R $, respectively, denotes the left- and right-hand function of Eq.~\eqref{eq:44},
namely
\begin{subequations}
  \begin{align}
    P_L \left( r \right)
    & =
      \frac{d^2}{dr^2}
      P \left( r \right)
      -
      \frac{\kappa \left( \kappa + 1 \right)}{r^2}
      P \left( r \right), \\
    P_R \left( r \right)
    & =
      \left\{
      -
      \frac{\left[ V \left( r \right) - E \right]^2}{c^2}
      +
      c^2
      \right\}
      P \left( r \right).
  \end{align}
\end{subequations}
Similarly, in Fig.~\ref{fig:figure4} (b),
$ Q_L $ and $ Q_R $, respectively, denotes the left- and right-hand function of Eq.~\eqref{eq:45},
namely
\begin{subequations}
  \begin{align}
    Q_L \left( r \right)
    & =
      \frac{d^2}{dr^2}
      Q \left( r \right)
      -
      \frac{\kappa \left( \kappa - 1 \right)}{r^2}
      Q \left( r \right), \\
    Q_R \left( r \right)
    & =
      \left\{
      -
      \frac{\left[ V \left( r \right) - E \right]^2}{c^2}
      +
      c^2
      \right\}
      Q \left( r \right).
  \end{align}
\end{subequations}
According to the second set of equations [Eqs.~\eqref{eq:44} and \eqref{eq:45}], 
$ P_L $ should be identical to $ P_R $, and $ Q_L $ should be identical to $ Q_R $.
Nevertheless, as evidenced in Fig.~\ref{fig:figure4},
neither the large- nor small-component wave functions satisfy the second set of equations.
This finding simultaneously provides crucial insights for resolving the anomalous rapid oscillations in the wave functions.
%


\section{The mono-kinetically-balanced generalized pseudospectral method}

\label{sec:GPM}
\par
Although the radial Dirac equation \eqref{eq14} involves only first-order differential operation on the wave function, the calculation of certain physical quantities still requires the wave function to be second-order differentiable.
In the GPS method, Eq.~\eqref{eq39-4} is used for the first-order differential operator.
Correspondingly, the first-order differentiation in the first set of equations to be satisfied
[Eqs.~\eqref{eq:41} and \eqref{eq:42}]
is also performed using Eq.~\eqref{eq39-4}.
Consequently, even the anomalously oscillating wave functions, such as those shown in Fig.~\ref{fig:figure2}, can satisfy this set of equations well.
However, since the second-order differential operator is not employed in the GPS scheme,
the obtained wave functions are not necessarily second-order differentiable
and accordingly satisfy the second set of equations.
Therefore, incorporating the second set of equations to be satisfied into the radial Dirac equation framework may potentially resolve these oscillation issues.
\par 
According to Eq.~\eqref{eq30}, the GPS method can be regarded as a special basis-set expansion method.
The basis functions for its two components can be expressed as
\begin{equation}
  \label{eq:46}
  u_i
  =
  \begin{cases}
    \begin{pmatrix}
      \pi_i \left( r \right) \\
      0
    \end{pmatrix}
    & \text{($ 1 \le i \le n $)}, \\
    \begin{pmatrix}
      0 \\
      \pi_{i - n} \left( r \right)
    \end{pmatrix}
    & \text{($ n + 1 \le i \le 2n $)}.
  \end{cases}
\end{equation}
Therefore, the large- and small-component wave functions can be respectively expressed as
\begin{subequations}
  \begin{align}
    P \left( r \right)
    & =
      \sum_i
      a_i
      \pi_i \left( r \right),
      \label{eq:47} \\
    Q \left( r \right)
    & =
      \sum_i
      b_i
      \pi_i \left( r \right),
      \label{eq:47_1}
  \end{align}
\end{subequations}
which must strictly satisfy the first set of equations [Eqs.~\eqref{eq:41} and \eqref{eq:42}],
\begin{subequations}
  \begin{align} 
    \sum_i
    a_i \pi_i \left( r \right)
    & =
      \frac{c}{V \left( r \right) + c^2 - E}
      \left(
      \frac{d}{dr}
      -
      \frac{\kappa}{r}
      \right)
      \sum_i
      b_i \pi_i \left( r \right), \\
    \sum_i
    b_i \pi_i \left( r \right)
    & =
      \frac{c}{c^2 + E - V \left( r \right)}
      \left(
      \frac{d}{dr}
      +
      \frac{\kappa}{r}
      \right)
      \sum_i
      a_i \pi_i \left( r \right).
      \label{eq:47_2}
  \end{align}
\end{subequations}
To incorporate the second set of equations to be satisfied [Eqs.~\eqref{eq:44} and \eqref{eq:45}],
taking Eq.~\eqref{eq:44} as an example,
the small-component wave function $ Q \left( r \right) $ must be expressed as
\begin{align}
  Q \left( r \right)
  & =
    \frac{c}{c^2 + E - V \left( r \right)}
    \left(
    \frac{d}{dr}
    +
    \frac{\kappa}{r}
    \right)
    P \left( r \right)
    \notag \\
  & =
    \frac{c}{c^2 + E - V \left( r \right)}
    \left( 
    \frac{d}{dr}
    +
    \frac{\kappa}{r}
    \right)
    \sum_i
    a_i \pi_i \left( r \right)
    \notag \\
  & =
    \frac{c}{c^2 + E - V \left( r \right)}
    \sum_i
    a_i
    \left(
    \frac{d}{dr}
    +
    \frac{\kappa}{r}
    \right)
    \pi_i \left( r \right).
    \label{eq:48}
\end{align}
Substituting Eqs.~\eqref{eq:47} and \eqref{eq:48} into Eq.~\eqref{eq:41},
one obtains 
\begin{align} 
  \sum_i
  a_i \pi_i \left( r \right)
  & = 
    \frac{c}{V \left( r \right) + c^2 - E}
    \left(
    \frac{d}{dr}
    -
    \frac{\kappa}{r}
    \right)
    \frac{c}{c^2 + E - V \left( r \right)}
    \sum_i
    a_i
    \left(
    \frac{d}{dr}
    +
    \frac{\kappa}{r}
    \right)
    \pi_i \left( r \right) 
    \notag \\
  & =
    \frac{c^2}{c^4 - \left[ E - V \left( r \right) \right]^2}
    \sum_i
    a_i
    \left[
    \frac{d^2}{dr^2}
    \pi_i \left( r \right)
    -
    \frac{\kappa \left( \kappa + 1 \right)}{r^2}
    \pi_i \left( r \right)
    \right],
    \label{eq:48_1}
\end{align}
which is mathematically equivalent to substituting Eq.~\eqref{eq:47} alone into Eq.~\eqref{eq:44}.
In the non-relativistic limit, Eq.~\eqref{eq:48} reduces to
\begin{equation}
  \label{eq:49}
  Q \left( r \right)
  =
  \frac{1}{2c}
  \sum_i
  a_i
  \left(
    \frac{d}{dr}
    +
    \frac{\kappa}{r}
  \right)
  \pi_i \left( r \right).
\end{equation}
This consequently yields the basis functions employed in the mono-kinetically-balanced (MKB) condition as referenced in Ref.~\cite{Igarashi2007,Jiao2021}
\begin{equation}
  \label{eq:50}
  u'_i
  =
  \begin{cases}
    \begin{pmatrix}
      \pi_i \left( r \right) \\
      0
    \end{pmatrix}
    & \text{($ 1 \le i \le n $)}, \\
    \begin{pmatrix}
      0
      \\
      D^{+} \pi_{i - n} \left( r \right)
    \end{pmatrix}
    & \text{($ n + 1 \le i \le 2n $)},
  \end{cases}
\end{equation}
where the operator $ D^{+} $ reads
\begin{equation}
  \label{eq:51}
  D^{+}
  =
  \frac{1}{2c}
  \left(
    \frac{d}{dr}
    +
    \frac{\kappa}{r}
  \right).
\end{equation}
The detailed implementation of the mono-kinetically-balanced condition in the GPS method can be found in Ref.~\cite{Jiao2021}.
Here, we present only the resulting MKB radial Dirac equation
\begin{equation}
  \label{eq:52}
  \begin{pmatrix}
    V \left( r \right) + c^2 & -c \left( \frac{d}{dr} - \frac{\kappa}{r} \right) \\
    c \left( \frac{d}{dr} + \frac{\kappa}{r} \right) & V \left( r \right) - c^2
  \end{pmatrix}
  \begin{pmatrix}
    1 & 0 \\
    0 & D^{+}
  \end{pmatrix}
  \begin{pmatrix}
    P_{n \kappa} \left( r \right) \\
    Q_{n \kappa} \left( r \right)
  \end{pmatrix}
  =
  E
  \begin{pmatrix}
    1 & 0 \\
    0 & D^{+}
  \end{pmatrix}
  \begin{pmatrix}
    P_{n \kappa} \left( r \right) \\
    Q_{n \kappa} \left( r \right)
  \end{pmatrix}
\end{equation}
and the new coupled radial equation in terms of $ x $ can be expressed as
\begin{equation}
  \label{eq:56}
  h'_{\urm{D}} \left( x \right)
  \phi' \left( x \right)
  =
  E
  O' \left( x \right)
  \phi' \left( x \right),
\end{equation}
where 
\begin{align}
  h'_{\urm{D}} \left( x \right)
  & =
    \begin{pmatrix}
      V \left( x \right) + c^2 & \frac{\kappa \left( \kappa + 1 \right)}{2 f^2 \left( x \right)} \\
      c \frac{\kappa}{f \left( x \right)} & \frac{V - c^2}{2c} \frac{\kappa}{f \left( x \right)}
    \end{pmatrix}
                                            +
                                            \begin{pmatrix}
                                              0 & 0 \\
                                              c & \frac{V - c^2}{2c}
                                            \end{pmatrix}
                                                  \frac{1}{\sqrt{f' \left( x \right)}}
                                                  \frac{d}{dx}
                                                  \frac{1}{\sqrt{f' \left( x \right)}}
                                                  \notag \\
  & \quad
    +
    \begin{pmatrix}
      0 & - 1/2 \\
      0 & 0
    \end{pmatrix}
          \frac{1}{f' \left( x \right)}
          \frac{d^2}{dx^2}
          \frac{1}{f' \left( x \right)} 
          \label{eq:57}
\end{align}
and
\begin{equation}
  \label{eq:58}
  O' \left( x \right)
  =
  \begin{pmatrix}
    1 & 0 \\
    0 & \frac{1}{2c} \frac{\kappa}{f \left( x \right)}
  \end{pmatrix}
  +
  \begin{pmatrix}
    0 & 0 \\
    0 & \frac{1}{2c}
  \end{pmatrix}
  \frac{1}{\sqrt{f' \left( x \right)}}
  \frac{d}{dx}
  \frac{1}{\sqrt{f' \left( x \right)}} .
\end{equation}
The second-order derivative of a function $ u \left( x \right) $ at collocation points can be approximated by~\cite{Canuto2006,ZhuLin2020}
\begin{equation}
  \label{eq:59}
  \left.
    \frac{d^2}{dx^2}
    u \left( x \right)
  \right|_{x =x_i}
  =
  \sum_{j = 0}^N
  \left( d_2 \right)_{ij}
  \frac{P_N \left( x_i \right)}{P_N \left( x_j \right)}
  u \left( x_j \right),   
\end{equation}
where
\begin{equation}
  \label{eq:60}
  \left( d_2 \right)_{ij}
  =
  \begin{cases}
    - \frac{2}{\left( x_i - x_j \right)^2}
    & \text{($ i\ne j $, $ i\in \left[ 1, N - 1 \right] $, $ j \in \left[ 0, N \right] $)}, \\
    - \frac{N \left( N + 1 \right)}{3 \left( 1 - x_i^2 \right)}
    & \text{($ i = j \in \left[ 1, N - 1 \right] $)}, \\
    \frac{N \left( N + 1 \right) \left( N^2 + N - 2 \right)}{24}
    & \text{($ i = j = 0 $ or $ N $)}, \\
    \frac{N \left( N + 1 \right) \left( 1 + x_j \right) - 4}{2 \left( 1 + x_j \right)^2}
    & \text{($ i = 0 $, $ j \in \left[ 1, N \right] $)}, \\
    \frac{N \left( N + 1 \right) \left( 1 - x_j \right) - 4}{2 \left( 1 - x_j \right)^2}
    & \text{($ i = N $, $ j \in \left[ 0, N - 1 \right] $}).
  \end{cases}  
\end{equation}
Therefore, the associated eigenvalue problem is expressed as
\begin{equation}
  \label{eq:61}
  \left[ \mathbf{h}'_{\urm{D}} \right]
  \left[ \mathbf{\phi}' \right]
  =
  \left[ \mathbf{E} \right]
  \left[ \mathbf{O}' \right]
  \left[ \mathbf{\phi}' \right],
\end{equation}
where the elements of matrices $ \mathbf{h}'_{\urm{D}} $ and $ \mathbf{O}' $ are
\begin{align}
  \left( h'_{\urm{D}} \right)_{ij}
  & = 
    \begin{pmatrix}
      V \left( f \left( x_i \right) \right) + c^2
      & \frac{\kappa \left( \kappa + 1 \right)}{2 f^2 \left( x_i \right)} \\
      c \frac{\kappa}{f \left( x_i \right)}
      & \frac{V \left( f \left( x_i \right) \right) - c^2}{2c} \frac{\kappa}{f \left( x_i \right)}
    \end{pmatrix}
        \delta_{ij}
        +
        \begin{pmatrix}
          0 & 0 \\
          c & \frac{V \left( f \left( x_i \right) \right) - c^2}{2c}
        \end{pmatrix}
              \frac{1}{\sqrt{f' \left( x_i \right)}}
              \left( d_1 \right)_{ij}
              \frac{1}{\sqrt{f' \left( x_j \right)}}
              \notag \\
  & \quad
    +
    \begin{pmatrix}
      0 & - 1/2 \\
      0 & 0
    \end{pmatrix}
          \frac{1}{f' \left( x_i \right)}
          \left( d_2 \right)_{ij}
          \frac{1}{f' \left( x_j \right)} 
          \label{eq:62}
\end{align}
and 
\begin{equation}
  \label{eq:63}
  \left( O' \right)_{ij}
  =
  \begin{pmatrix}
    1 & 0 \\
    0 & \frac{1}{2c} \frac{\kappa}{f \left( x_i \right)}
  \end{pmatrix}
  \delta_{ij}
  +
  \begin{pmatrix}
    0 & 0 \\
    0 & \frac{1}{2c}
  \end{pmatrix}
  \frac{1}{\sqrt{f' \left( x_i \right)}}
  \left( d_1 \right)_{ij}
  \frac{1}{\sqrt{f' \left( x_j \right)}},
\end{equation}
respectively.
\begin{figure}
  \centering
  \includegraphics[width=1.0\textwidth]{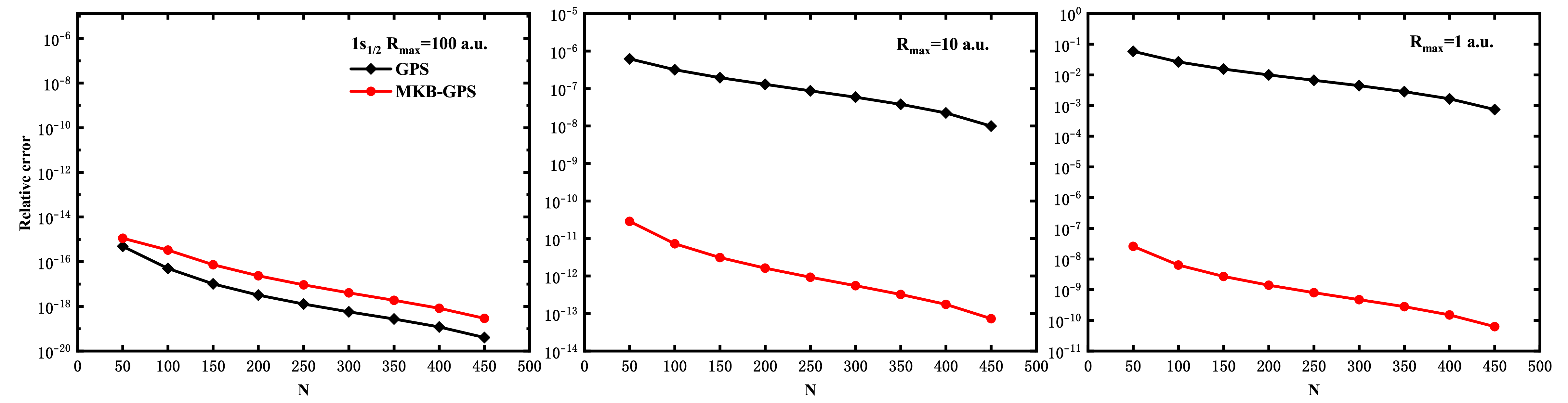}
  \caption{Relative errors in the $ 1s_{1/2} $ state energies of the confined hydrogen atom calculated by the GPS and MKB-GPS methods with different confinement radii $ R_{\urm{max}} $.}
  \label{fig:figure5}
\end{figure}
\begin{figure}
  \centering
  \includegraphics[width=1.0\textwidth]{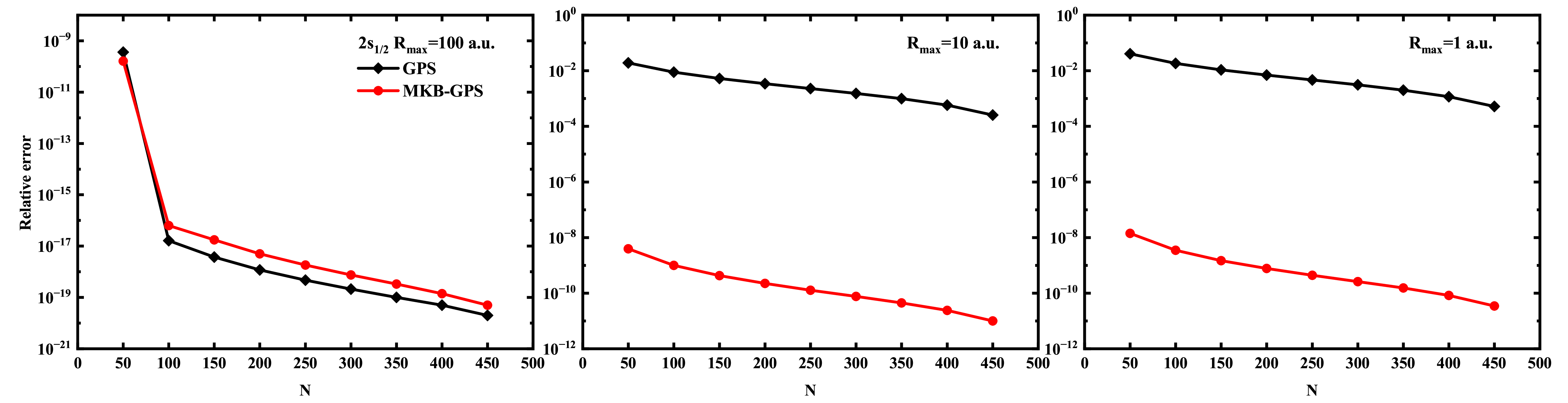}
  \caption{Relative errors in the $ 2s_{1/2} $ state energies of the confined hydrogen atom calculated by the GPS and MKB-GPS methods with different confinement radii $ R_{\urm{max}} $.}
  \label{fig:figure6}
\end{figure}

\begin{figure}
    \centering
    \includegraphics[width=1\linewidth]{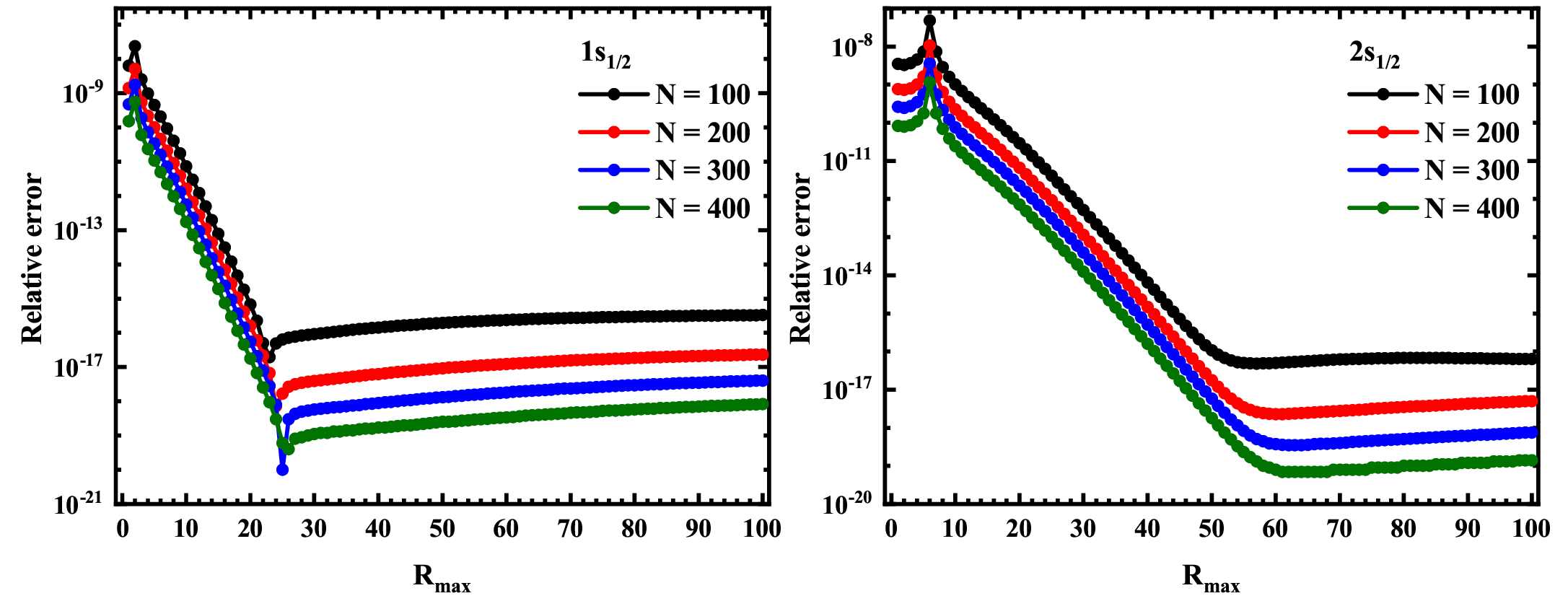}
    \caption{Relative errors of the bound-state energies of the hydrogen atom calculated by MKB-GPS method with the different $ N $.}
    \label{fig:figure6-1}
\end{figure}
\par
\begin{figure}
  \centering
  \includegraphics[width=1.0\textwidth]{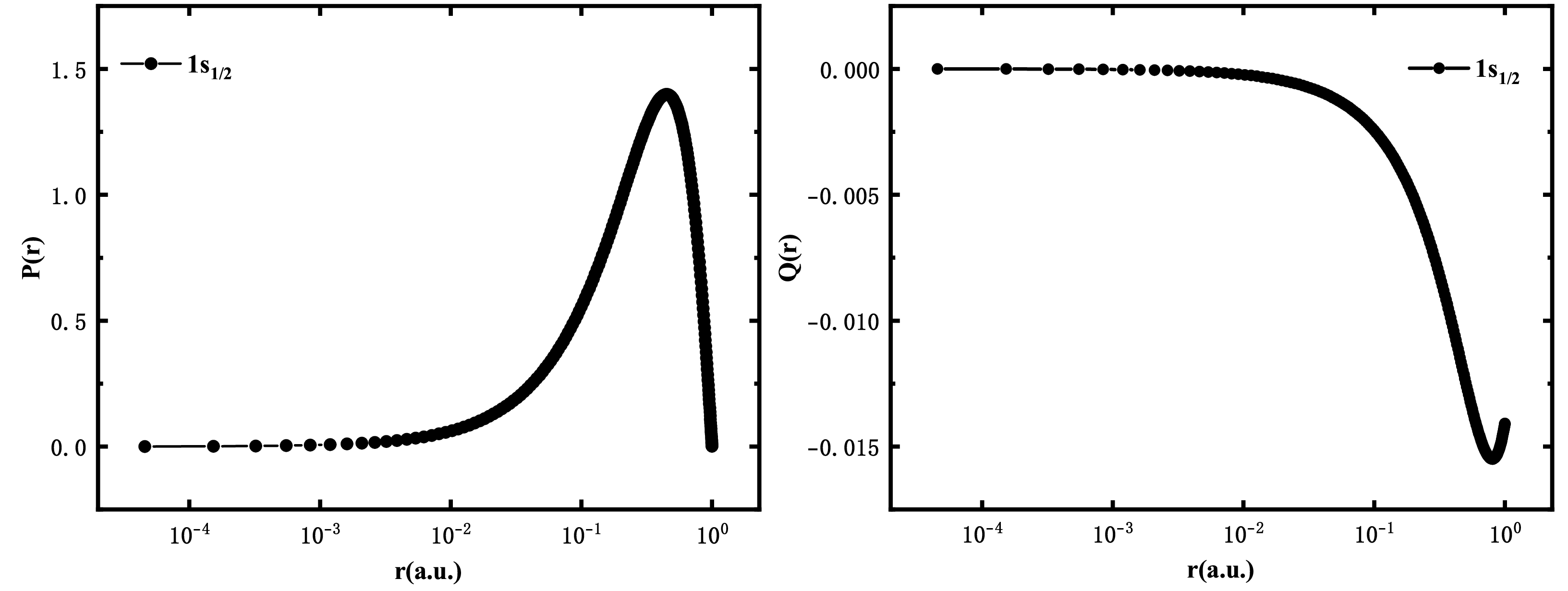}
  \caption{The radial wave function of the $ 1s_{1/2} $ state of the hydrogen atom calculated by MKB-GPS method with the confinement radius $ R_{\urm{max}} = 1 \, \mathrm{a.u} $.}
  \label{fig:figure7}
\end{figure}
\begin{figure}[h!]
  \centering
  \includegraphics[width=1.0\textwidth]{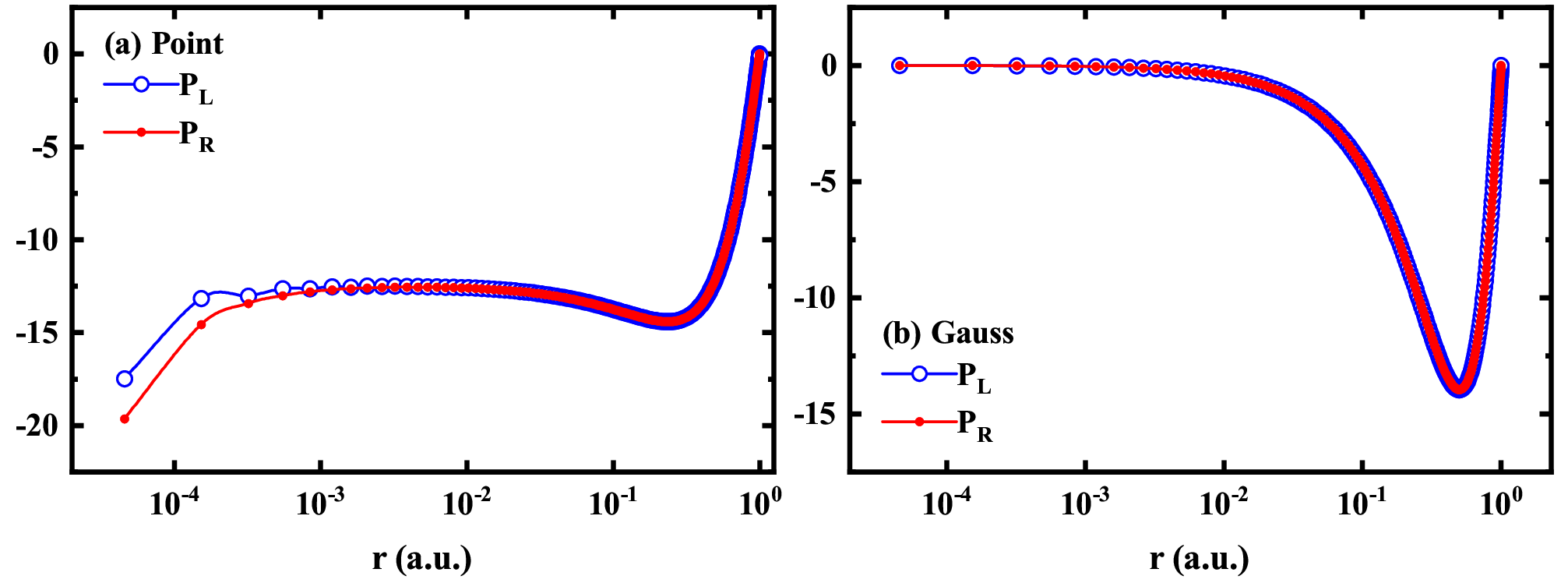}
  \caption{Functions on the left- and right-hand sides of the second set of equations to be satisfied.}
  \label{fig:figure9}
\end{figure}

Figures~\ref{fig:figure5} and \ref{fig:figure6} present the relative errors in the energies of the $ 1s_{1/2} $ and $ 2s_{1/2} $ states of the confined hydrogen atom,
calculated using the GPS and MKB-GPS methods for different confinement radii $ R_{\urm{max}} $.
The reference energies for the black and red lines are obtained using the GPS and MKB-GPS methods, respectively, with $ N = 500 $. 
As shown in both figures, when the confinement radius is $ R_{\urm{max}} = 100 \, \mathrm{a.u.} $,
the difference between the relative errors computed by the GPS and MKB-GPS methods is negligible,
on the order of $ 10^{-17} $.
However, as the confinement radius decreases (e.g., $ R_{\urm{max}} = 10 \, \mathrm{a.u.} $ and $ 1 \, \mathrm{a.u.} $),
the MKB-GPS method reduces relative errors by about six orders of magnitude compared to the GPS method,
indicating significantly improved computational accuracy.
As can be seen from Fig.~\ref{fig:figure6-1}, although the relative error still begins to increase when the confinement radius $ R_{\urm{max}} $ decreases to a certain value,
the magnitude of the increase is much smaller compared to that shown in Fig.~\ref{fig:figure1_1}.
Moreover, over the range of confinement radii considered in Fig.~\ref{fig:figure6-1}, the relative error remains acceptable.
Figure~\ref{fig:figure7} demonstrates that the MKB-GPS method successfully resolves the oscillation issues in the small-component wave functions, yielding smooth and continuous wave functions.
Figure~\ref{fig:figure9} shows that the wave functions obtained using the MKB-GPS method satisfy the second set of equations well.
The slight discrepancy near the origin in Fig.~\ref{fig:figure9} (a) arises because the Coulomb potential diverges at $ r = 0 $.
Indeed, if the Gaussian-charge distribution model---whose potential remains finite at $ r = 0 $---is used instead,
the corresponding result shown in Fig.~\ref{fig:figure9} (b)
exhibits the exact equality between the left- and right-hand sides of the second set of equations at the origin.
%


\section{Conclusion}
\label{sec:con}
\par
Numerical solutions of the Dirac equation under confining potentials using the GPS method exhibit deteriorating convergence of energy eigenvalues and emerging high-frequency oscillations in wave functions when the confinement radius decreases below a critical range.
Unlike the spurious states that appear in the GPS method, these situations occur in all the obtained energy eigenvalues and wave functions.
Starting from the radial Dirac equation, two sets of equations are derived.
The first set involves first-order differentiation of the wave function, while the second set involves second-order differentiation.
These two sets of equations are employed to examine the wave function. It is found that the rapidly oscillating wave functions satisfy the first set of equations but fail to satisfy the second set.
Based on this inference, the oscillation problem originates from the formula used for first-order differentiation in the GPS method. Furthermore, if the oscillating wave functions could be made to satisfy the second set of equations, this type of problem might be resolved.
Motivated by this idea, an attempt is made to introduce the kinetically-balanced condition into the GPS method, i.e., the MKB-GPS method.
Originally, the kinetically-balanced condition was proposed to address the spurious states encountered in the GPS method. In the present context, the constraints imposed by this condition on the wave function enable it to satisfy the second set of equations.
The final results demonstrate that the MKB-GPS method successfully addresses this issue, yielding converged energy eigenvalues as well as smooth and continuous wave functions.
In this work, two sets of equations that the wave function must satisfy are used to analyze the problems encountered in the numerical solution.
Moreover, this is the first time that the MKB-GPS method is applied to confined potentials.
Its effectiveness is verified even for small confinement radii, where good results are obtained.
These findings significantly expand the applicability of kinetic balance approaches and establish a robust framework for future investigations of relativistic quantum systems in confined potentials.


\section*{Declarations}




\subsection*{Acknowledgements} 
\par
Dengshan Liu, Huihui Xie, Pengxiang Du and Jian Li acknowledge the National Natural Science Foundation of China  (Grants Nos.~12475119 and 12447101) and the China Postdoctoral Science Foundation, No.~2025M783390.
T.~N.~acknolwedges
the JSPS Grant-in-Aid under Grant Nos.~JP24K17057,
JP25H00402,
JP25H01558,
JP25K01003,
and
JP25KJ0405,
and
JST COI-NEXT Grant No.~JPMJPF2221.


\medskip

\printbibliography

\end{document}